\documentclass[onecolumn, accepted=2019-02-23]{quantumarticle}

\usepackage{graphics}
\usepackage{hyperref}
\usepackage{color}
\usepackage{listings}
\usepackage{epsf}
\usepackage{epsfig}
\usepackage{rotating}
\usepackage{subfigure}
\usepackage{amsmath}
\usepackage{amssymb}
\usepackage{amsfonts}
\usepackage{mathrsfs}
\usepackage{latexsym}
\usepackage{authblk}
\usepackage{textpos}
\usepackage{bbm}
\usepackage{dsfont}
\usepackage{epstopdf}
\usepackage{amsthm}
\usepackage{braket}
\usepackage{xcolor}
\usepackage{url}
\hypersetup{
    colorlinks,
    linkcolor={black!50!black},
    citecolor={blue!50!black},
    urlcolor={blue!80!black}
}

\numberwithin{equation}{section}

\newcommand{\HH}{\mathcal{H}}

\newcommand{\Tr}{\textrm{Tr}}

\begin{document}

\title{Determining a local Hamiltonian from a single eigenstate}

\author[1,2]{Xiao-Liang Qi}
\author[1]{Daniel Ranard}
\affil[1]{\small \em  Stanford Institute for Theoretical Physics, Stanford University, Stanford CA 94305 USA}
\affil[2]{\small \em  Institute for Advanced Study, Princeton NJ 08540 USA}
\date{}

\maketitle

\begin{abstract}

We ask whether the knowledge of a single eigenstate of a local Hamiltonian is sufficient to uniquely determine the Hamiltonian.  We present evidence that the answer is ``yes" for generic local Hamiltonians, given either the ground state or an excited eigenstate.  In fact, knowing only the two-point equal-time correlation functions of local observables with respect to the eigenstate should generically be sufficient to exactly recover the Hamiltonian for finite-size systems, with numerical algorithms that run in a time that is polynomial in the system size.  We also investigate the large-system limit, the sensitivity of the reconstruction to error, and the case when correlation functions are only known for observables on a fixed sub-region.  Numerical demonstrations support the results for finite one-dimensional spin chains (though caution must be taken when extrapolating to infinite-size systems in higher dimensions). For the purpose of our analysis, we define the ``$k$-correlation spectrum" of a state, which reveals properties of local correlations in the state and may be of independent interest.

\noindent
\let\thefootnote\relax\footnotetext{\hspace{-0.75cm}{\tt \, xlqi$@$stanford.edu, \tt dranard$@$stanford.edu}}
\end{abstract}

\newpage
\tableofcontents
\newpage

\section{Introduction}

Can one determine the Hamiltonian of a system from knowledge of a single eigenstate? Without any restrictions on the form of the Hamiltonian, the answer is clearly ``no," because infinitely many Hermitian operators share the same eigenstate. However, most Hermitian operators are not physically sensible Hamiltonians. If we place strong enough restrictions on the class of Hamiltonians (due to some pre-existing knowledge about the system), it becomes plausible that a single eigenstate does determine the Hamiltonian. For instance, given a single-particle Hamiltonian of the form $H = p^2 + V(x)$, the unknown function $V(x)$ can be determined up to a constant by any eigenstate wavefunction $\psi(x)$.  Another case was studied by Ref.\ \cite{Grover}, demonstrating that a many-body Hamiltonian may be recovered from the knowledge of a single excited eigenstate (with nonzero energy density) whenever the system satisfies the eigenstate thermalization hypothesis (ETH) \cite{Deutsch, Srednicki}.  

In this paper, we address this question for a broad class of many-body Hamiltonians: local lattice Hamiltonians with finite-range interactions.  More precisely, we consider many-body systems with a tensor-factorization of the Hilbert space $\mathcal{H}=\otimes _x\mathcal{H}_x$, where each factor $\mathcal{H}_x$ is a finite-dimensional Hilbert space associated with a site $x$. Given the tensor factorization, we can define range-$k$ local operators as operators that only act nontrivially on $k$ spatially contiguous sites. For Hamiltonians that are linear combinations of range-$k$ local operators, we argue that almost all such Hamiltonians may be uniquely recovered from the knowledge of a single eigenstate.  We first provide an intuitive argument based on parameter counting, and then we develop a more constructive approach by defining a quantity we call the $k$-correlation spectrum. Our methods can be generalized straightforwardly to certain other classes of Hamiltonians, such as $k$-local Hamiltonians (defined as linear combinations of terms acting on $k$ sites that may not be spatially contiguous).

The intuitive parameter-counting argument is presented in Section \ref{sec:dimCounting}, which we first sketch here. Define a map from local Hamiltonians to states, mapping each Hamiltonian to its $i$'th eigenstate, for some fixed $i$. Because the dimension of the vector space of local Hamiltonians is polynomial in system size (that is, the number of lattice sites), while the dimension of Hilbert space is exponential in system size, the image of this map lies in a relatively low-dimensional submanifold of the Hilbert space.  We might therefore expect the map to generically have non-singular derivative -- which, for this particular map, implies injectivity.

The above argument is only suggestive, and there do exist mulitple Hamiltonians with the same eigenstate.  For instance, if the Hamiltonian has zero couplings between lattice sites, the eigenstates will be separable (non-entangled), with many Hamiltonians sharing that eigenstate.  More generally, if the Hamiltonian commutes with a local conserved quantity $O$, then the modified Hamiltonian $H \to H+\lambda O$ must share all non-degenerate eigenstates with $H$. However, these cases are not generic.  Granting an assumption explained in Section \ref{sec:genericity}, we then show that such counter-examples form a measure-zero set in the space of local Hamiltonians on finite-size systems.

In Section \ref{sec:correlationMatrix} we develop a constructive approach to obtain the Hamiltonian from one eigenstate.  We introduce the matrix of connected correlations of (spatially) local observables of range $k$, which we call the $k$-correlation matrix $M^{(v)}$ associated to any state $v$ in the Hilbert space.  Specifically, we define
\begin{equation}\label{eq:correlationMatrix}
M^{(v)}_{ij} = \frac{1}{2} \bra{v}\{ L_i, L_j \}\ket{v} -\bra{v} L_i \ket{v}\bra{v} L_j \ket{v}
\end{equation}
where $\{\cdot,\cdot\}$ denotes the anti-commutator and $L_i \in \text{Herm}(\mathcal{H})$ are operators that form an orthonormal basis for the space spanned by operators local to spatially contiguous regions of size $k$. 

We call the eigenvalue spectrum of this matrix the $k$-correlation spectrum, and it may be calculated for any state, without reference to a Hamiltonian.  This spectrum reveals information about the correlations and entanglement properties of the state, and it is invariant under local unitary operations. We will observe that the number of zeros in the correlation spectrum is the dimension of the subspace of traceless local Hamiltonians that have $v$ as an eigenstate.  (For convenience, we will generally restrict our analysis to traceless Hamiltonians, in order to disregard the component of the Hamiltonian proportional to the identity, which only introduces a constant shift in energy.)  Thus $v$ is an eigenstate of some traceless, nonzero local Hamiltonian if and only if the correlation spectrum has at least one zero, and such a Hamiltonian is unique (up to an overall scaling\footnote{We will subsequently omit the caveat that the reconstructed Hamiltonian is only unique up to scaling and adding multiples of the identity.}) if and only if the correlation spectrum has exactly one zero.

After introducing the correlation matrix, we calculate correlation spectra numerically for eigenstates of one-dimensional spin chains.  For randomly generated local Hamiltonians on up to $n=12$ qubits, both in the disordered case and the translation-invariant case, we are able to numerically recover the Hamiltonian from the knowledge of any single eigenstate.  In fact, granting certain assumptions supported by numerical evidence, we prove analytically that almost all Hamiltonians may be uniquely and exactly determined by any single eigenstate.

For practical purposes, it's important to ask how sensitive the reconstruction is to error.  If the given eigenstate has some error relative to the actual eigenstate, the smallest nonzero eigenvalue of the correlation spectrum bounds the sensitivity of the reconstructed Hamiltonian to this error.  Moreover, if the eigenstate  is only known on some sub-region of the entire system, the Hamiltonian on this region may still be approximately recovered in some cases.  The question of approximate recovery is addressed in Section \ref{sec:sensitivity}.

The correlation spectrum of a state is interesting in its own right.  We investigate how the correlation spectrum of an eigenstate varies with the energy of the eigenstate. For translation-invariant states, the spectrum may be organized into bands (or rather curves) parameterized by momentum.  We find that ground states of gapped, translation-invariant systems exhibit a certain band gap in the correlation spectrum. 

Finally, these results may have direct application for experimental setups.  For instance, consider a translation-invariant spin system in the laboratory, where the Hamiltonian and even the ground state may not be known, but where the correlation functions of local observables in the ground state may be measured in some fixed region.  Then the Hamiltonian could be approximated using the correlation matrix, as discussed in Section \ref{sec:subregion}.

\section{Reconstructing the Hamiltonian using the correlation matrix} 

\subsection{Setup and dimension counting}\label{sec:dimCounting}
First we present a heuristic argument that a single eigenstate of a local Hamiltonian uniquely determines the Hamiltonian.  Consider a quantum system on a finite lattice of spatial dimension $d$ with $n$ sites, where the dimension of the Hilbert space at each site is $m$, for total Hilbert space $\mathcal{H}$ of dimension $N = m^n$. By a ``range-$k$ local Hamiltonian," we mean a Hamiltonian that may be written as a sum of interactions of $k$ spatially contiguous sites.  Throughout the paper, we will only directly consider finite systems. 

The vector space of all Hamiltonians has real dimension $N^2=m^{2n}$, but the space of local Hamiltonians is much smaller, with dimension of order $nm^{2k^d}$. Most importantly, the space of local Hamiltonians has dimension that scales linearly with system size $n$, unlike the space of states, which scales exponentially with $n$. This observation provides the basis for the ability to recover the Hamiltonian from an eigenstate, which then only requires recovering a small number of parameters (linear in $n$), starting with a large number of parameters (exponential in $n$).  We will see that actually, only the correlations of local observables are required to recover the Hamiltonian, and these correlations are described by a number of parameters quadratic in $n$. 

The exact dimension counting is easy to see in one dimension.  For simplicity, consider a qubit spin chain (i.e.\ spatial dimension $d=1$ with local Hilbert space dimension $m=2$) with nearest-neighbor interactions (range $k=2$) and periodic boundary conditions.  The most general traceless Hamiltonian with these properties may be written using Pauli matrices as
\begin{equation}\label{eq:QubitHam}
H=\sum_{i=1}^n H_i = \sum_{i=1}^n \sum_{a=0}^3 \sum_{b=1}^3 c^i_{ab} \sigma^i_a \sigma^{i+1}_b
\end{equation}
with $12n$ real parameters $c^i_{ab}$, where $a=0,1,2,3$ and $b=1,2,3$.  Denote this space of local Hamiltonians $LH \subset \text{Herm}(\HH)$, with real dimension $\dim(LH)=12n$.  Equivalently, $LH$ is the vector space spanned by local operators of range $k=2$. The set of states in $\HH$ that are the unique eigenstate of a local Hamiltonian must then be contained within a submanifold $\mathcal{M} \subset \HH$ of real dimension at most $12n$, whereas the Hilbert space $\HH$ has real dimension $2N=2^{n+1}$. For systems of sufficiently large size $n$, $\dim(LH) \ll \dim(\HH)$.  Hence almost all states in $\HH$ are not the unique eigenstate of any local Hamiltonian. 
A similar argument indicates that almost all states cannot be in the degenerate eigenspace of a local Hamiltonian, provided the degeneracy is not exponential in system size.  (Moreover, one may explicitly construct states that are guaranteed to be far from any eigenstate of any local Hamiltonian \cite{Osborne}.)

Now we ask, for states that \textit{are} eigenstates of some local Hamiltonian, is that Hamiltonian unique?  Consider the mapping from $LH$ to $\HH$ taking a local Hamiltonian to its $i$'th eigenstate for some fixed $i$.  This map goes from a relatively small space to a much larger space, so we might expect it to have non-singular derivative at most points.  Because every fiber of this map  (i.e.\ the preimage of a point in the image) is a linear subspace, a non-singular derivative implies the fiber contains exactly one point, or rather the Hamiltonian can be recovered from the eigenstate.

Of course, dimension counting alone does not guarantee that the map has a non-singular derivative. We analyze uniqueness more rigorously and explicitly in the following subsections.

\subsection{Defining the correlation matrix}\label{sec:correlationMatrix}

Given a state $v \in \HH$, how can one determine whether it is the eigenstate of some local Hamiltonian, and in particular whether the Hamiltonian may be uniquely recovered?  First, note that a Hermitian operator $O$ has eigenstate $v$ if and only if 
\begin{equation}
\langle O^2 \rangle_v - \langle O \rangle_v^2=0,
\end{equation} i.e.\ the variance or ``fluctuation" of $O$ with respect to $v$ is zero.  The above condition directly follows if $v$ is an eigenstate of $O$.  Conversely, to see that the above condition implies $v$ is an eigenstate of $O$, decompose the vector $O|v\rangle$ into a term proportional to $|v\rangle$ and orthogonal to $|v\rangle$, i.e.\ 
\begin{equation}
O|v\rangle = \alpha |v\rangle + \alpha_\perp |v_\perp \rangle,
\end{equation}
for some normalized $|v_\perp \rangle$. Then 
\begin{equation} \label{eq:sizePerpendicular}
\langle O^2 \rangle_v - \langle O \rangle_v^2=|\alpha_\perp|^2,
\end{equation}
and if the above quantity is zero, then $\alpha_\perp=0$ and hence $v$ must be an eigenvector of $O$.

We will re-express this condition using the correlation matrix $M^{(v)}$ of Equation \ref{eq:correlationMatrix},
\begin{equation*} 
M^{(v)}_{ij} = \frac{1}{2}\langle\{L_i, L_j\} \rangle_v -\langle L_i \rangle_v \langle L_j \rangle_v,
\end{equation*}
where the expectations are taken with respect to state $v$, and $L_i \in \textrm{Herm}(\HH)$ are operators that form a basis for $LH \subset \textrm{Herm}(\HH)$, the space of range-$k$ local Hamiltonians. Equivalently, $LH$ is spanned by operators local to spatially contiguous subsystems of size $k$; we occasionally refer to these sums of local operators as local operators themselves, so that ``local Hamiltonian" and ``[Hermitian] local operator" have identical meaning.  While we have allowed the freedom to choose any basis $\{L_i\}$ of $LH$, it is useful to choose a basis such that each operator $L_i$ is local to a particular region of size $k$, rather than being a linear combination of such operators.  For instance, the $\{L_i\}$ could always be chosen as a subset of products of (generalized) Pauli operators.  With such a choice, the entries of the correlation matrix will actually be two-point connected correlations of operators of range $k$, rather than sums of such correlations.

We choose $\{L_i\}$ orthonormal in the Hilbert-Schmidt inner product, with a normalization convention $\|\mathds{1}\| \equiv \frac{1}{N}\Tr(\mathds{1}^2)=1$.  Then the condition that operator $O = w_i L_i$ has eigenstate $v$ may be written
\begin{align}  \label{eq:corrMatFluctuation}
\langle O^2 \rangle_v - \langle O \rangle_v^2 & = w^T M^{(v)} w \\
& = \bra{O}M^{(v)}\ket{O} \nonumber \\
& = 0. \nonumber
\end{align}
The second line is simply an alternative notation that will be useful later, treating the operator $O$ as a vector in the vector space $LH$ of local Hamiltonians.  

Therefore one can think of $M^{(v)}$ as a symmetric bilinear form on the vector space $LH$.  The bilinear form simply maps operators to their correlation with respect to $v$.  That is, for operators $O_1$ and $O_2$,
\begin{equation}
\bra{O_1}M^{(v)}\ket{O_2} = \langle O_1 O_2 \rangle_v - \langle O_1 \rangle_v \langle O_2 \rangle_v.
\end{equation}

Equivalently, we can see $M^{(v)}$ as a (super-)operator on $LH$.  By the condition of Equation \ref{eq:corrMatFluctuation}, the kernel of $M^{(v)}$ is precisely the space of local Hamiltonians with eigenstate $v$.  Recalling Equation \ref{eq:sizePerpendicular}, we can think of $\bra{O}M^{(v)}\ket{O}=|\alpha_\perp|^2$ as measuring the degree to which $O$ fails to have $v$ as an eigenstate.

For an eigenstate $v$ of a local Hamiltonian, the kernel of $M^{(v)}$ must have dimension at least one. If $\dim \ker M^{(v)}$ is exactly one, there is no other Hamiltonian with $v$ as an eigenstate, so that the Hamiltonian is uniquely recovered from $v$. The correlation matrix then provides a simple construction of the Hamiltonian: For a given state $v$, we first calculate the correlation matrix $M^{(v)}$ (which has relatively small dimension, quadratic in system size).  Or in some situations, we might start with $M^{(v)}$ directly, without ever knowing $v$. Then we diagonalize $M^{(v)}$ to look for its zero eigenvalue, which takes a polynomial time $O(n^3)$.  (The correlation matrix has dimension $O(n)$, because it includes correlations of only spatially local observables.) If there is no zero eigenvalue, we conclude that $v$ is not an eigenstate of any local Hamiltonian. If there is a unique zero eigenvalue, with the eigenvector $w_i$, then $H=\sum_iw_iL_i$ is the unique local Hamiltonian with state $v$ as an eigenvector. If $M^{(v)}$ has more than one zero eigenvalue, the Hamiltonian cannot be uniquely determined by $v$.  We call the spectrum of $M^{(v)}$ the ``correlation spectrum" of $v$, which depends on the subspace $LH$ but is evidently independent of the basis $\{L_i\}$. 

\begin{figure}[t]
  \centering
    \includegraphics[width=1.0\textwidth]{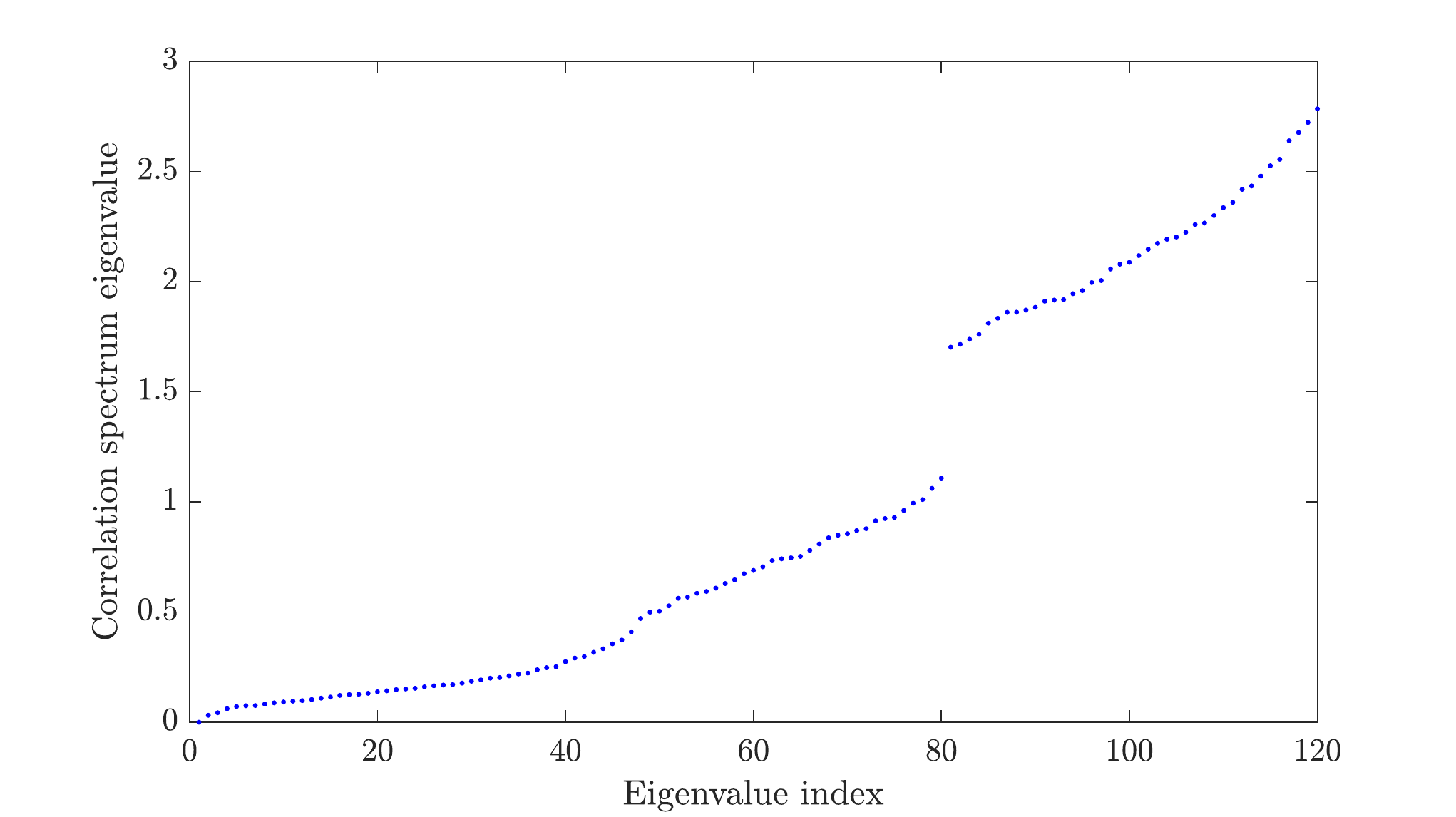}
     \caption{The correlation spectrum is shown for the ground state of a 10-qubit spin chain Hamiltonian with randomly generated coefficients (non-translation-invariant).  Although the reconstructions were tested for up to $n=12$ qubits, an example with $n=10$ is shown here for visual clarity (to see how the eigenvalues are separated).}
\label{fig:gsCorr}
\end{figure}

As an example, we numerically study spin chain Hamiltonians as in Equation \ref{eq:QubitHam} with i.i.d.\ random Gaussian couplings $c^i_{ab}$ acting on 12 qubits.  The correlation spectrum of each eigenstate was found to have a single zero, implying the reconstructed Hamiltonian is unique.  Indeed, the reconstructed Hamiltonian consistently matches the original to within error of about $\theta \approx 10^{-10}$, with $\theta$ defined in Equation \ref{eq:theta}.  The correlation spectrum for the ground state of such a Hamiltonian is shown in Figure \ref{fig:gsCorr}; note the single zero at the bottom left corner.

Given only the two-point correlation functions of range-$k$ observables for an eigenstate, the reconstructed Hamiltonian could then be used to determine the eigenstate itself.  (Each ``two-point" correlation probes up to $2k$ sites.)  In this way, one reconstructs the eigenstate from its local correlations. It is already well-known that (non-degenerate) eigenstates of local Hamiltonians are uniquely determined by their local correlations \cite{Zeng, Cramer, Swingle}.  The results here take this conclusion a step further: one recovers not only the eigenstate but also the Hamiltonian.  (Reference \cite{Cramer} constructs ``parent Hamiltonians'' for states using only local expectation values, but these parent Hamiltonians are not generally range $k$, so they do not recover the original range-$k$ Hamiltonian.)  

In the case of ground states, the state is actually already determined by the expectation values of single local observables (rather than the two-point correlation functions needed to recover excited eigenstates).  Thus for nearest neighbor translation-invariant systems, the ground state is uniquely determined by the reduced density matrix (RDM) on any two adjacent sites.  Ref. \cite{Verstraete} discusses how this 2-body RDM reveals properties of the ground state.

The correlation matrix may also be written
\begin{equation}\label{eq:corrMatMixedComm}
\bar{M}^{(\rho)}_{ij} = \frac12 \Tr([\rho,L_i]^+[\rho,L_j]),
\end{equation}
where $\rho=\ket{v}\bra{v}$, which extends the definition of the correlation to mixed states.  (The bar is used to distinguish $\bar{M}^{(\rho)}$ from the $M^{(\rho)}$ of Equation \ref{eq:corrMatMixedExp}, which offers a \textit{different} generalization of the correlation matrix to mixed states.)
In this form, the correlation matrix shows a resemblance to the quantity
\begin{equation} 
M^{(H)}_{ij} = \frac12  \Tr([H,L_i][H,L_j]),
\end{equation}
studied with slight modifications in \cite{Huse} and \cite{Vidal}.  While the eigen-operators of $M^{(v)}$ with the smallest eigenvalues are the operators that have the least fluctuation with respect to $\rho$, the eigen-operators of $M^{(H)}_{ij}$ with the smallest eigenvalues are the operators that change most slowly in time.

\begin{sloppypar}
\subsection{Showing the reconstructed Hamiltonian is generically unique}\label{sec:genericity}
\end{sloppypar}

For some $H \in LH$ with eigenstate $v$, we now ask whether to expect that the correlation spectrum of $v$ has only one zero, allowing unique recovery of $H$.  Denote the correlation spectrum of $v$ as 
\begin{equation}
\textrm{spectrum}(M^{(v)})=\{\lambda^{(v)}_i\}_{i=1}^N
\end{equation}
with eigenvalues ordered from smallest to largest, $\lambda^{(v)}_1 \leq  \cdots \leq \lambda^{(v)}_N$.  Then for any eigenstate $v$ of a local Hamiltonian, $\lambda^{(v)}_1=0$, and the Hamiltonian may be reconstructed if and only if $\lambda^{(v)}_2>0$.  We want to show that for almost all $H \in LH$, for any eigenstate $v$ of $H$, the only Hamiltonian that has eigenstate $v$ is $H$. That is, we want to show that almost all $H \in LH$ may be reconstructed from any one of their eigenstates.  We therefore prove the following result, which \textit{assumes there exists some $H_0$ for which the reconstruction is unique}:

\bigskip

\noindent{\bf Theorem.} For a given Hilbert space and tensor factorization into $n$ sites, with a given subspace of local operators $LH$, let us assume there exists a single Hamiltonian $H_0$ with non-degenerate spectrum, such that the $i$'th eigenstate $v$ of $H_0$ allows unique reconstruction of $H_0$. In other words, we assume there exists some $H_0$ for which the $i$'th eigenstate has correlation spectrum with exactly one zero.  Given this assumption, the set of Hamiltonians in $LH$ that cannot be reconstructed from their $i$'th eigenstate has measure zero.

\noindent{\bf Proof.}  Let $X \subset LH$ be the set of local Hamiltonians with non-degenerate spectrum.  Then $H_0 \in X$, and $X$ is a connected open set with measure zero complement.  Consider the $i$'th eigenstate $v$ of $H_0$.  By definition, $\lambda^{(v)}_1 =0$ for all $H \in LH$, and by assumption, $\lambda^{(v)}_2>0$ for $H_0$.  The argument will essentially run as follows: we consider $\lambda^{(v)}_2$ as a function of $H$, and this function will be analytic (subject to subtleties concerning whether $\lambda^{(v)}_2$ is a degenerate eigenvalue).  Crucially, the set of zeros of an analytic function has measure zero, unless the function itself is zero. So if $\lambda^{(v)}_2$ is nonzero for some point $H_0 \in X$, then $\lambda^{(v)}_2$ must be nonzero almost everywhere in $X$, implying almost all $H \in LH$ may be reconstructed from their $i$'th eigenstate. To avoid the subleties mentioned above, the argument below does not directly consider $\lambda^{(v)}_2$ as a function of $H$, but it is similar in spirit.

Because $\lambda^{(v)}_1 =0$ and $\lambda^{(v)}_2>0$ for $H_0$, $M^{(v)}$ has nullity $1$, or rank $N-1$.  Note that a matrix has rank at least $r$ if and only if all of its $r$-by-$r$ matrix minors have nonzero determinant.  Hence all $(N-1)$-by-$(N-1)$ matrix minors of $M^{(v)}$ have nonzero determinant.  Moreover, the determinant is a polynomial of the matrix entries of $M^{(v)}$.  In turn,  $M^{(v)}$ has entries whose real and imaginary parts are real polynomials of the real and imaginary parts of $v$, the $i$'th eigenstate of $H$.  Consider the map $f_i : X \to \mathbb{P}(\HH)$ sending a Hamiltonian to the ray of its $i$'th eigenstate, where $\mathbb{P}(\HH)$ is the projective Hilbert space.  The map $f_i$ is analytic.  Then the determinants of the $(N-1)$-by-$(N-1)$ matrix minors of $M^{(v)}$ depend analytically on $H$, and so the determinants must be nonzero almost everywhere.  Hence $M^{(v)}$ has rank $N-1$ almost everywhere, and so $\lambda^{(v)}_2>0$ almost everywhere, and $H$ may be reconstructed from $v$ for almost all $H \in LH$. $\blacksquare$

\bigskip

For a lattice of some fixed finite size, any single example of a local Hamiltonian that may be recovered from its eigenstate would then serve as rigorous proof that almost all Hamiltonians on this lattice are recoverable; this is the power of analyticity.  Such examples were found numerically  for spin chains of size up to $n=12$, as illustrated in Figure \ref{fig:gsCorr}. This numerical evidence effectively proves that for chains of $n=12$ qubits, almost all Hamiltonians are recoverable from their eigenstates.  

The observations in Section \ref{sec:largeLimit} suggest how results for specific systems of finite size might be used to prove statements about generic systems of arbitrarily large size.  However, in the infinite-size limit, the analysis is more complicated, especially because the subset $X$ of non-degenerate local Hamiltonians may no longer be dense nor connected, due to the existence of topological phases in $d \geq 2$ spatial dimensions.  Thus an example of a Hamiltonian recoverable from its ground state would only show that almost all Hamiltonians in that same quantum phase are also recoverable from their ground states.

In the case of a translation-invariant Hamiltonian, we may consider the same Hamiltonian extended to systems of different size.   As mentioned in Section \ref{sec:largeLimit}, the ability to recover a Hamiltonian from its ground state for systems of a certain size may then guarantee the ability to recover the same Hamiltonian from its ground state on all larger systems, assuming an energy gap. 

Finally, let us note a relationship between the above discussion and the ``quantum geometric tensor" introduced by \cite{Zanardi}.  Here, we have considered the map from Hamiltonian to eigenstate.  In the case of ground states, this map was studied in a different context when thinking about quantum phase transitions.  The quantum geometric tensor measures how the ground state changes with respect to perturbations in the Hamiltonian, and it may be calculated explicitly with perturbation theory.

\bigskip
\subsection{Translation-invariant systems and banded correlation spectra}

Translation invariance simplifies the the analysis in several ways.  For translation-invariant systems of qudits, a range-$k$ local Hamiltonian is specified by a finite number of parameters that is independent of system size.  For a periodic system, the translation invariance also allows the correlation spectrum to be divided into bands,\footnote{We say ``band" in analogy with a banded energy spectrum, but each band is actually a one-dimensional curve, unlike the thickened band that appears in band theory.} much in the same way that translation-invariant Hamiltonians have a banded energy spectrum, with different eigenvalues within a single band corresponding to different momenta.  

Consider a translation-invariant state $v$, the eigenstate of some translation-invariant local Hamiltonian.  More generally, $v$ might be an eigenstate of nonzero momentum, but we will assume $v$ has zero momentum to simplify the discussion.  For a translation-invariant state, it is helpful to consider the correlation matrix $M^{(v)}$ in a momentum basis, where it will become block diagonal.  More specifically, first consider a basis 
\begin{equation}
\{L_{x\alpha}\}
\end{equation} for the space of range-$k$ local Hamiltonians $LH$, where the index $x=1,...,n$ labels the site, $L_{x\alpha}$ has support within sites $x$ to $x+k$-1, and $\alpha=1,...,S/n$ labels all range-$k$ local operators at fixed position, with $S=\dim(LH)$.
The correlation matrix for a translation-invariant state may then be written
\begin{align}  \label{eq:opCorrBasis}
M^{(v)}_{x\alpha,y\beta}  & =  \frac{1}{2} \langle 
\{ L_{x\alpha}, L_{y\beta} \} \rangle_v -\langle L_{x\alpha} \rangle_v \langle L_{y\beta}\rangle_v\\
& \equiv  f_{\alpha\beta}(|x-y|)   \nonumber
\end{align}
where $\{f_{\alpha\beta}(r)\}_{\alpha,\beta=1}^{S/n}$ denote the two-point correlation functions of operators labeled by $\alpha$ and $\beta$, with separation $r \in \{0,...,n-1\}$.

Denote the Fourier-transformed basis as
$\{\widetilde L_{q\alpha}\}$, for $\alpha = 1,..,S/n$ and momentum $q=2 \pi j/n$ with $j \in \{0,...,n-1\}$.  Specifically, define
\begin{equation}
 \widetilde L_{q\alpha} = \frac{1}{\sqrt{n}} \sum_{x=1}^n L_{x \alpha} e^{iqx}.
\end{equation}
For instance, the operators $\widetilde L_{q \alpha}$ for $q=0$ span the space of translation-invariant local Hamiltonians.  For a translation-invariant eigenstate $v$ with zero momentum, it is straightforward to show that the correlation matrix in the momentum basis becomes block diagonal,
\begin{align}
\widetilde M^{(v)}_{p\alpha,q\beta} & =  \frac{1}{2} \langle 
\{ \widetilde L_{p\alpha}, \widetilde L_{q\beta} \} \rangle_v -\langle \widetilde L_{q\alpha} \rangle_v \langle\widetilde L_{p\beta} \rangle_v \\
& = \delta_{pq} \widetilde M^{(v)}_{p\alpha,q\beta} \\ \nonumber
& =  \delta_{pq} \widetilde f_{\alpha\beta}(p) \nonumber
\end{align}
where $\widetilde f_{\alpha\beta}(p)$ are the Fourier-transformed correlation functions
\begin{equation}
\widetilde  f_{\alpha\beta}(p) = \frac{1}{\sqrt{n}} \sum_x f_{\alpha \beta}(j)e^{i p x}
\end{equation}
For fixed $p$, we have the matrix $\widetilde M^{(v)}_{p\alpha, p\beta} = f_{\alpha \beta}(p)$, with indices $\alpha,\beta$ taking on values $1$ to $S/n$.  It may be diagonalized to yield the eigenvalues $\{\lambda_\alpha^q\}_{\alpha=1}^{S/n}$,
\begin{equation} \label{eq:qSpec}
\{\lambda_\alpha^q\}_{\alpha=1}^{S/n} = \textrm{spectrum}(M^{(v)}_{p\alpha, p\beta}).
\end{equation}

In the $n \to \infty$ large-system limit, $p \in [0,2\pi)$ becomes a continuous parameter.  Because the Fourier transform of an exponentially decaying function is smooth, the Fourier space correlation functions $\widetilde f_{\alpha\beta}(p)$ will depend smoothly on $p$ if the position space correlation functions $f_{\alpha\beta}(|x-y|)$ are exponentially decaying in distance.  Such states with exponentially decaying correlations include gapped ground states, eigenstates of many-body localized systems, and eigenstates of systems satisfying ETH, with the last case subject to caveats discussed at the end of the section.  When the correlation matrix entries depend smoothly on $q$, the set of eigenvalues will also depend smoothly on $q$, so that the eigenvalues form momentum bands.  The number of bands is $S/n=\dim(LH)/n$, independent of the number of lattice sites $n$. For example, $S/n=12$ for qubit spin chains. For a fixed translation-invariant Hamiltonian on system of finite size $n$, the eigenvalues in each momentum band should become continuous as $n \to \infty$ and tend toward some limiting curves.

While the full structure of the bands is helpful for characterizing properties of the state, only the zero-momentum sector of the correlation spectrum is relevant to the reconstruction of a translation-invariant Hamiltonian from its eigenstate.   That is, one only needs to consider the part of the correlation matrix $\widetilde M^{(v)}_{p\alpha,q\beta}$ at $p,q=0$, corresponding to the correlation matrix of translation-invariant observables.  The reconstruction of the Hamiltonian is then possible if and only if the eigenvalue spectrum of the translation-invariant correlation matrix has exactly one zero.  The eigenvector in the kernel of the translation-invariant correlation matrix specifies the parameters of the recovered Hamiltonian.

\begin{figure}[t] 
  \centering
    \includegraphics[width=0.9\textwidth]{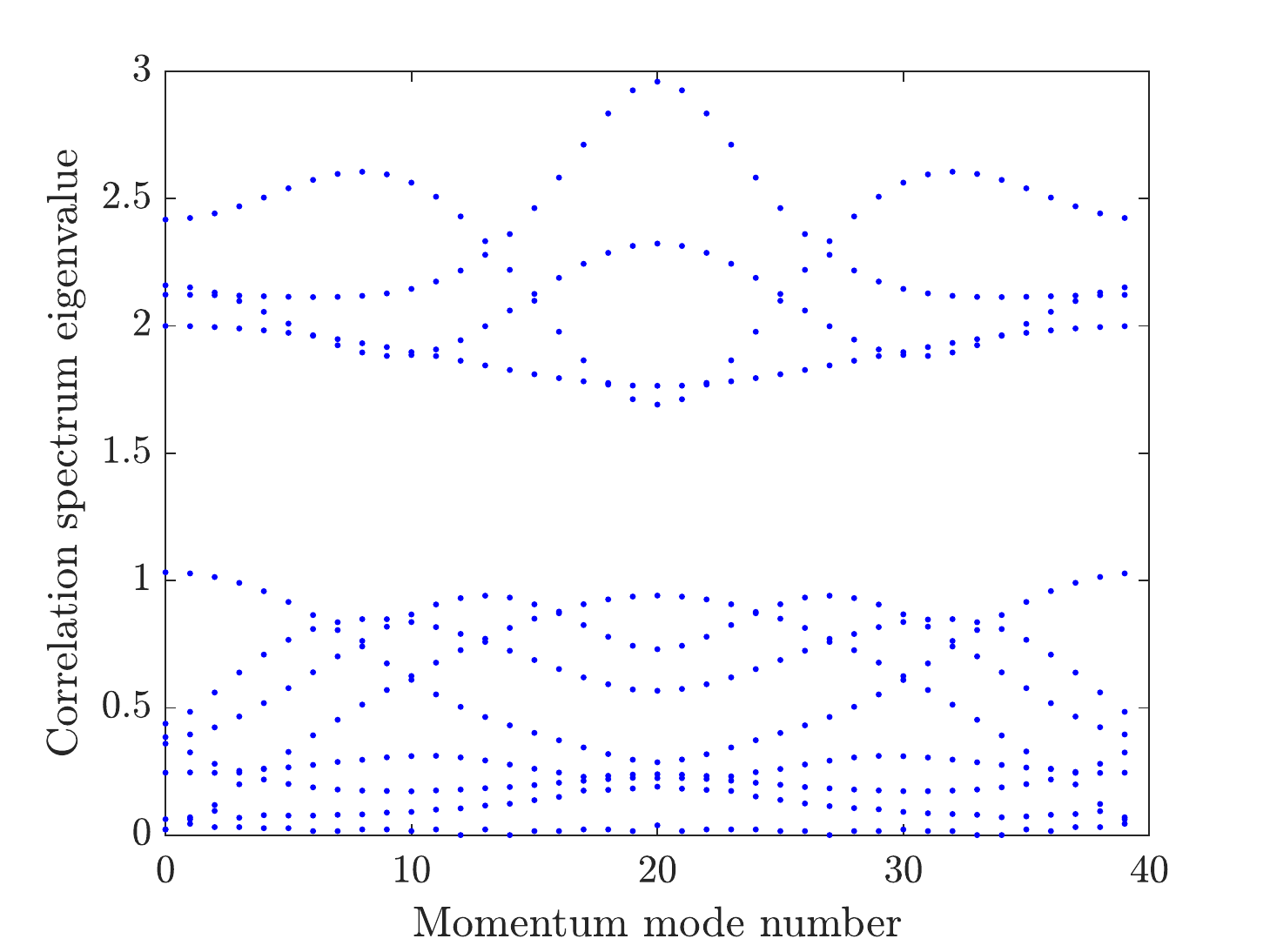}
     \caption{The correlation spectrum is shown for the ground state of a 40-qubit translation-invariant spin chain Hamiltonian with randomly generated coefficients.   The ``momentum mode number" refers to integer $j=0,...,n-1$ in the momentum $q=2\pi j/n$.  Unlike in other examples, the ground state here was found approximately using matrix product state methods rather than exact diagonalization, so that a larger system with more discretized momentum values could be used, to better showcase the structure of the bands.}
\label{fig:gsCorrBand}
\end{figure}

An example of the banded correlation spectrum is shown in Figure \ref{fig:gsCorrBand} for the ground state of a 40-qubit translation-invariant spin chain with randomly generated couplings.  The correlation spectrum has a single zero, which occurs in the lowest band at $q=0$.  The corresponding eigen-operator of the correlation matrix yields the reconstructed Hamiltonian, and the single zero in the correlation spectrum implies that the reconstructed Hamiltonian is unique.  As expected, the reconstructed Hamiltonian numerically matches the original, randomly generated Hamiltonian. Note that the bands are approximately smooth, due to the exponential decay of correlations in a gapped ground state.  The example shown is qualitatively typical for gapped, zero-momentum ground states of randomly generated, translation-invariant spin chain Hamiltonians.

Again, the $q=0$ eigenvalue of the lowest band corresponds to the Hamiltonian itself, $H = \sum_x H_x$.  Meanwhile, the eigenvalues at $q \neq 0$ in the lowest band are associated with eigen-operators that approximately correspond to the energy density times a nonzero phase factor, i.e.\ $\sum_x H_x e^{iqx}$.  For arbitrarily large systems, as the band becomes continuous, the correlation spectrum has many eigenvalues arbitrarily close to zero.  As discussed further in Section \ref{sec:sensitivity}, small nonzero eigenvalues in the correlation spectrum imply that the reconstructed Hamiltonian is highly sensitive to numerical error, or to error in the known eigenstate.  This sensitivity is easy to understand: small nonzero eigenvalues in the correlation spectrum correspond to candidate Hamiltonians that are orthogonal to the true Hamiltonian but for which the given state is still an approximate eigenstate.  

When the Hamiltonian is known to be translation-invariant, the very small eigenvalues in the lowest band discussed above do not necessarily imply an error-prone reconstruction, because eigenvalues may be ignored by restricting attention to the correlation matrix at $q=0$.  However, even if the Hamiltonian were not known to be translation-invariant, there is a sense in which these small eigenvalues would not pose a problem for reconstruction: although the operator $\sum_x H_x e^{iqx}$ is orthogonal to the Hamiltonian, it approximately matches the Hamiltonian up to an overall phase when truncated to regions of size less than $q^{-1}$.  

When reconstructing the Hamiltonian from exponentially decaying correlation functions, this sort of error is inevitable: the correlations in a local region only aid in reconstructing the Hamiltonian on that region, and only up to an overall phase; on length scales much larger than the correlation length, one cannot guarantee that the Hamiltonians reconstructed on each region are stitched together with the correct phase to yield the correct total Hamiltonian.

The preceding discussion becomes more important in the case of non-translation-invariant Hamiltonians, for which operators like $\sum_x H_x e^{iqx}$ for small $q$ will again lead to small nonzero eigenvalues in the correlation spectrum.  In that case, one can only hope to recover the Hamiltonian up to this slowly varying phase factor.

\begin{figure}[t] 
  \centering
    \includegraphics[width=1.0\textwidth]{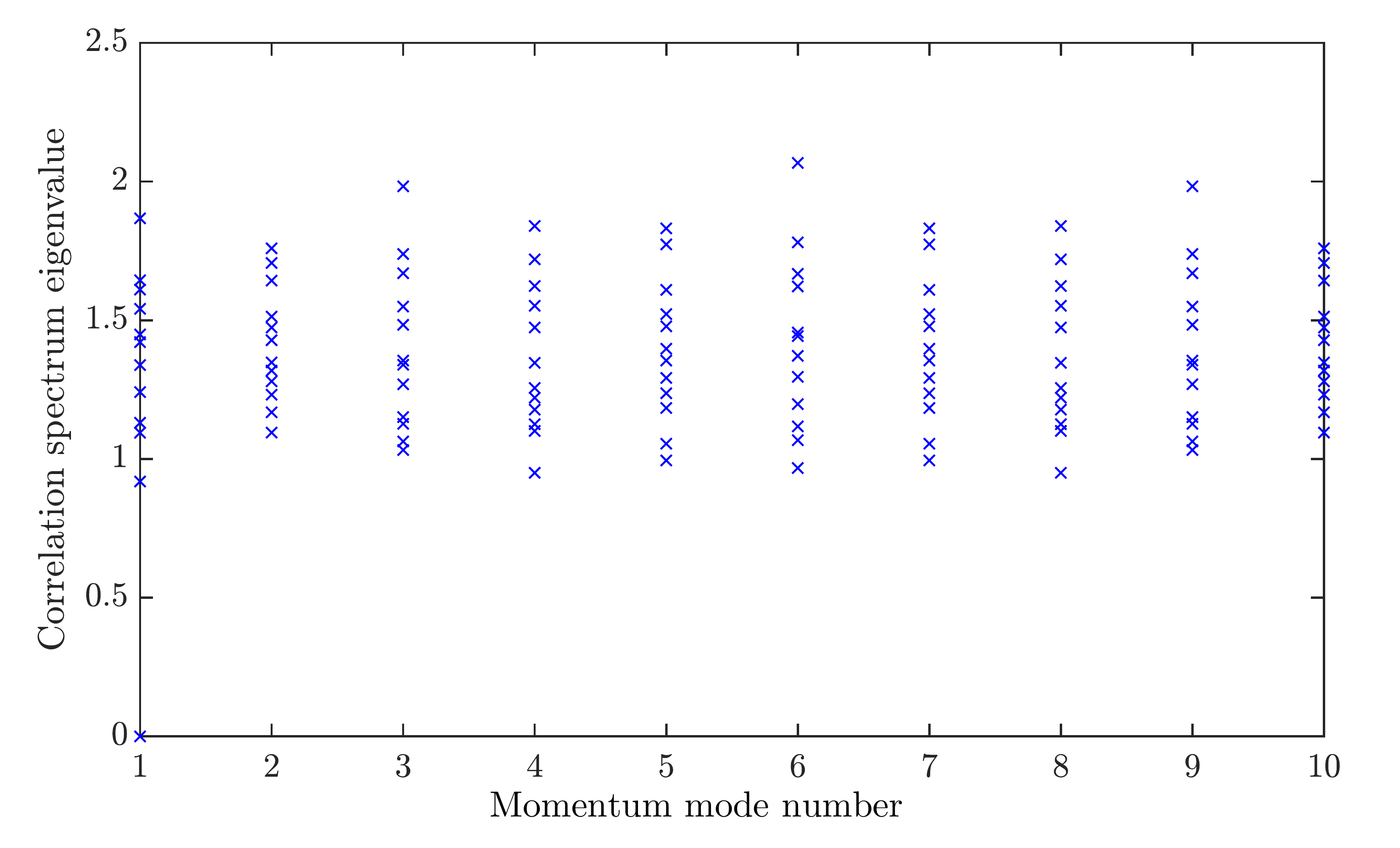}
     \caption{The correlation spectrum is shown for a mid-spectrum excited state of a 10-qubit translation-invariant spin chain Hamiltonian with randomly generated coefficients.  The ``momentum mode number" refers to integer $j=0,...,n-1$ in the momentum $q=2\pi j/n$.}
\label{fig:esCorrBand}
\end{figure}

Meanwhile, Figure \ref{fig:esCorrBand} shows the correlation spectrum for a mid-spectrum excited state of the same translation-invariant Hamiltonian with randomly generated coefficients (though on a system with fewer qubits, due to computational constraints).  The correlation spectrum again has a single zero (in the lower left corner), yielding an accurate reconstruction of the Hamiltonian.  However, note the lowest band appears to have a discontinuity at $q=0$, implying that the two-point correlations of the energy terms $H_x$ in the Hamiltonian $H = \sum_i H_i$ must not decay exponentially (otherwise the band would be smooth).  The example shown is qualitatively typical of mid-spectrum excited states of randomly generated, translation-invariant spin chain Hamiltonians.

In fact, we can argue that the correlation functions of energy in an eigenstate that satisfies ETH \textit{never} decay exponentially, thereby yielding the discontinuity in the lowest band of the correlation spectrum.  The key fact is that for a state with exponentially decaying correlations, the total fluctuation (i.e.\ variance) of any operator $O = \sum_i O_i$ is approximately the sum of the fluctuation in each sub-region, where the total lattice has been partitioned into sub-regions of sufficient size.  That is, for states where correlations of $O_i$ decay exponentially,
\begin{equation}
\langle O^2 \rangle- \langle O \rangle ^2 \approx  \sum_A \langle O_A^2 \rangle - \langle O_A \rangle ^2 = \sum_A \sum_{i,j \in A} \langle O_i O_j \rangle - \langle O_i \rangle \langle  O_j \rangle  
\end{equation}
where the lattice has been partitioned into regions $A$ of length much greater than the correlation length of the state.  For an eigenstate that satisfies ETH on regions small compared to the total system size, the fluctuation of the Hamiltonian in any such region will be equal to the thermal fluctuation in that region, so the right-hand side of the above will be proportional to total volume. But the total fluctuation of the Hamiltonian will be zero, so the above equation cannot hold.  Hence the correlations of energy cannot decay exponentially for ETH eigenstates, at least on scales comparable to the total system size $n$; the correlations must be of order at least $\frac{1}{n}$. This nonzero, non-decaying correlation of energy density at large distances makes sense when considering energy conservation: measuring an extra unit of energy in one region implies measuring $\frac{1}{n}$ fewer units of energy elsewhere, on average. A related discussion appears in \cite{Grover}.

The correlation spectra shown in Figures \ref{fig:gsCorrBand} and \ref{fig:esCorrBand} and appear qualitatively typical of such randomly generated translation-invariant local Hamiltonians, at least when the ground state is gapped with momentum $q=0$. Notice that for the ground state in Figure \ref{fig:gsCorrBand}, the spectrum has a gap in momentum bands, between the lower 8 bands and the upper 4.  This gap appears typical.  On the other hand, the correlation spectrum in Figure \ref{fig:esCorrBand} for a mid-spectrum excited state shows no such gap.  Thus the gap in the correlation spectrum appears to distinguish ground states from excited states.   Excited states also have a larger smallest-nonzero-eigenvalue, allowing more robust reconstruction of the Hamiltonian.

Why is the gap between bands present in the correlation spectrum of the ground state?  No conclusive calculation is presented here.  However, it is illuminating to consider the correlation spectrum of a product state, such as the all-spin-up state $\prod|\uparrow\rangle$ in the qubit chain, which yields 8 flat bands at exactly $\lambda=0$ and 4 bands at $\lambda=1$, producing a gap similar to that of a general gapped ground state.  If a ground state is related to the product state by an adiabatic evolution, we expect the gap to stay open for a finite parameter range, since the correlation spectrum is a continuous function of the state. We numerically verified that if a sufficiently short quantum circuit is applied to the product state, the bands will change but the gap will persist.  (Rather than using a circuit, to preserve translation invariance one may instead consider evolution for a short time with a translation-invariant local Hamiltonian.)  The eigen-operators will only obtain slightly expanded spatial support after evolution via a short circuit, and these operators may be approximated with a truncation to subsystems of size $k=2$. Intuitively, we expect that the operators in the bands below the gap come from finite-size truncations of almost-local conserved quantities, but a precise relation is not known. 

\section{Approximate reconstruction}

\subsection{Reconstruction from data with error}\label{sec:sensitivity}

If the eigenstate is only known approximately -- if the given state has some ``error" relative to the true eigenstate -- then the reconstructed Hamiltonian will have some corresponding error relative to the true Hamiltonian.  Understanding the sensitivity of the reconstruction to error is important for several reasons.  First, the error analysis is necessary when extrapolating results about smaller systems to arbitrarily large systems, as discussed in Section \ref{sec:largeLimit}. If the sensitivity of the reconstruction to error were to diverge in the large-system limit, the Hamiltonian would not be robustly determined by an eigenstate for large systems, and the numerics performed for finite systems would be misleading.  The error analysis is also relevant for experimental applications, where it is possible that (1) the correlations of local observables are measured with error, (2) the state is not an exact eigenstate, and (3) measurements of correlations are only known for observables confined to a sub-region within the entire system.  We will find that the sensitivity to error is related to the correlation spectrum.

First we consider a situation in which the correlations are known for the whole system, but they are measured with some error. For some eigenstate $v$ of $H$, let the measured correlation matrix be
\begin{equation}
M'^{(v)} = M^{(v)} + \epsilon \Delta M^{(v)}.
\end{equation}
(If the true state of the system were not an exact eigenstate, the error could be considered in the same way.)  Assume that $M^{(v)}$ has exactly one zero eigenvalue, i.e.\ $v$ allows unique reconstruction of $H$.  Call the vector in the kernel $\ket{H}$, following the notation under Equation \ref{eq:corrMatFluctuation}, where the components of $\ket{H}$ correspond to the components of $H$ in the space $LH$ of local Hamiltonians.  In general $M'^{(v)}$ may have trivial kernel, but one still obtains a candidate reconstruction $\ket{H'}$ by taking a vector of smallest eigenvalue.  Take $\ket{H}$ and $\ket{H'}$ to have norm 1.  The error of $H'$ relative to $H$ may be quantified as 
\begin{equation} \label{eq:theta}
\theta = \cos^{-1}\| \braket{H | H'}\|, 
\end{equation}
i.e.\ the angle between the reconstructed Hamiltonians, or their distance in the space of projectivized Hamiltonians.  To check that $\theta$ is a sensible measure of error, one must understand its normalization and how it scales with system size.  Consider two translation-invariant Hamiltonians $H=\sum_{i=1}^n h_i$ and $H'=\sum_{i=1}^n h'_i$, for convenience normalized to $\|H\|=\|H'\|=1$ with $\|h_i\|,|h'_i\| \propto \frac{1}{\sqrt{n}}$.  Let the terms $h_i$ be chosen to be orthogonal for different $i$, and likewise for $h'_i$.  Then the angle $\theta$ between $h_i$ and $h'_i$ is also the angle $\theta$ between $H$ and $H'$, so $\theta$ is a measure of the distance between the individual terms of the Hamiltonians, and this quantity does not scale with the total system size for translation-invariant systems.

Calculating this quantity is an application of the perturbation theory of matrices and eigenvectors, familiar from quantum mechanics.  Mathematically, $\ket{H'}$ is the perturbed ``ground state" (eigenvector of lowest eigenvalue) of $M'^{(v)}$.  In particular,
\begin{equation}\label{eq:pertOfCorr}
\ket{H'} \propto \ket{H}+ \epsilon \sum_{i>1 }^S \frac{\bra{O_i}\Delta M^{(v)} \ket{H}}{\lambda_i-\lambda_1} \ket{O_i} + O(\epsilon^2)
\end{equation}
where $\ket{O_i}$ are the eigenvectors (or rather eigen-operators) of $M^{(v)}$, $\lambda_i$ are the eigenvalues in the correlation spectrum, and $\lambda_1=0$.  The above expression for $\ket{H'}$ is not normalized; accounting for normalization, one obtains the bound
\begin{align}
1-\| \braket{H | H'}\| \leq & \frac{\epsilon^2}{2} \left\|\sum_{i>1 }^s  \frac{\bra{O_i}\Delta M^{(v)} \ket{H}}{\lambda_i-\lambda_1} \right\| + O(\epsilon)^3 \\ 
\leq & \frac{\epsilon^2}{2} \frac{1}{\lambda_2^2} \| \Delta M^{(v)}\|^2+  O(\epsilon^4),
\end{align}
where $\lambda_2$ is the second-smallest eigenvalue of the correlation spectrum.  Using a small angle approximation,
\begin{equation}
\theta = \cos^{-1}\| \braket{H | H'}\|  \lesssim \epsilon \| \Delta M^{(v)}\| \frac{1}{\lambda_2}+ O(\epsilon^2).
\end{equation}

The above error bound considers only the first derivative of the error with respect to infintesimal perturbations of the $M^{(v)}$.  Rather than merely bound the infintesimal sensitivity of $H$ to $v$, one may also bound the error in $H$ due to non-infintesimal perturbations of $v$ or $M^{(v)}$, using the Davis-Kahan sin-theta theorem \cite{DavisKahan}. The theorem yields the exact bound
\begin{equation} 
\frac12\sin{2 \theta} \leq \epsilon  \frac{\| \Delta M^{(v)}\| }{\lambda_2}.
\end{equation}

Therefore the error of the reconstructed Hamiltonian is inversely proportional to the smallest nonzero eigenvalue $\lambda_2$ of the correlation spectrum.  

If the Hamiltonian is known to be translation-invariant, then we can restrict our attention to the correlation matrix of translation-invariant observables, $M^{(v)}_{0 \alpha, 0 \beta}$, whose dimension $S/n$ will be independent of system size, with spectrum $\lambda_i^q$ (see Equation \ref{eq:qSpec}).  The error of the reconstructed Hamiltonian is then inversely proportional to $\lambda_2^{q=0}$, the smallest nonzero eigenvalue of the correlation spectrum for zero momentum.  By the discussion of the preceding section, for an eigenstate with exponential decay of correlations, one expects $\lambda_2^{q=0}$ to have some finite limiting value in the large-system limit.  Then the sensitivity of the reconstruction would be upper bounded in the large-system limit, assuming $\lambda_2^{q=0}$ tends to a nonzero value.  

Finally, we can gain alternative perspective on the correlation spectrum and the sensitivity of the reconstruction in a more geometric way.  Given $H \in LH$ with eigenstate $v$, consider the subspace $O_v \subset \text{Herm}(\mathcal{H})$ of Hermitian operators that have $v$ as an eigenstate.  By definition, $H \in O_v$, and the intersection $LH \cap O_v \subset \text{Herm}(\mathcal{H})$ is precisely the space of traceless local Hamiltonians that have $v$ as an eigenstate, {\it i.e.} the kernel of the correlation matrix $M^{(v)}$. For the purpose of reconstruction, we are interested in $\dim(LH \cap O_v)$, which must have dimension at least one.  Generically, we might expect $\dim(LH \cap O_v)=1$, because $\dim(LH)+\dim(O_v)<\dim(\textrm{Herm}(\HH))$.  If the planes $LH$ and $O_v$ intersect at a small angle, then small errors in $v$ (and hence $O_v$) may generate large deviations of the intersection. The ``angle" between two hyperplanes is characterized by a list of angles called principal angles. A calculation in linear algebra reveals that the principal angles between subspaces $L$ and $O_v$ are actually related to the correlation spectrum by
\begin{equation}
\theta_i = \cos^{-1}\left(1-\lambda_i^2\right),
\end{equation}
where $\{\lambda_i\}_{i=1}^s$ is the correlation spectrum of $v$.  For small angles, the equation yields $\theta_i \approx 2\lambda_i$. We see that the smallest nonzero eigenvalues in the correlation spectrum provide a measure of the smallest angles between the subspaces $L$ and $O_v$, again indicating the sensitivity of the reconstruction of $H$.

\subsection{Reconstruction from data restricted to a sub-region}\label{sec:subregion}

Now we consider the case that the correlations are only known for observables local to a particular sub-region of the system.  In particular, consider a qubit spin chain of length $n$, where correlations are only known for a subsystem $A$ consisting of $m$ contiguous sites.  Assume the full system is in eigenstate $v$ of the Hamiltonian $H$, and let $\rho_A=\Tr_{\bar{A}}(\ket{v}\bra{v})$ be the reduced state of $v$ on $A$.  Using the notation of Equation \ref{eq:opCorrBasis}, we assume that we only know the entries of the correlation matrix $M^{(v)}_{x \alpha, y \beta}$ for $x,y \in A$.  This sub-matrix of the correlation matrix may also be expressed as $M^{(\rho_A)}$, where we define the correlation matrix for a general mixed state as 
\begin{equation} \label{eq:corrMatMixedExp}
M_{ij}^{(\rho)} =  \frac{1}{2}\Tr(\rho\{L_i, L_j\})- \Tr(\rho L_i) \Tr(\rho L_j).
\end{equation}

Can one recover $H$ from knowledge of $M^{(\rho_A)}$?  When the system is disordered (that is, not translation-invariant), one could only hope to recover $H_A$, the sum of terms in $H$ that are local to $A$.  When the system is translation-invariant, one could hope to recover $H$ in its entirety.  

For both the translation-invariant and disordered case, the possibility for recovery is very different for excited states at nonzero energy density than for ground states.  For an eigenstate $v$ at nonzero energy density, if the system satisfies the eigenstate thermalization hypothesis (ETH) -- which is thought to hold for generic non-integrable systems -- then $\rho_A$ is approximated by the thermal Gibbs state for $m \ll n$:
\begin{align}
\rho_A & \approx \frac{\Tr_{\bar{A}}(e^{-\beta H})}{\Tr(e^{-\beta H})} \\ \nonumber
& \approx \frac{e^{-\beta H_A}}{\Tr(e^{-\beta H_A})}
\end{align} 
where the inverse-temperature $\beta$ is defined by $\frac{dS(E)}{dE}|_{E=E_v} = \beta$, for eigenstate $v$ at energy $E_v$, with density of states $S(E)$.  Then 
\begin{equation}
\log(\rho_A) \propto H_A + const.,
\end{equation}
 so knowledge of $\rho_A$ is sufficient to approximately recover $H_A$.  The subtleties and implications of this procedure are discussed in detail in reference \cite{Grover}.

When the thermal Gibbs state $\rho \propto e^{-\beta H}$ is given, the Hamiltonian may be obtained by taking a logarithm.  However, this method of obtaining $H$ requires knowledge of $\rho$, not just the 2-point local correlation functions encoded by $M^{(\rho)}$.  The correlation matrix $M^{(\rho)}$ of the thermal state generally has no kernel, so the method used for recovery from eigenstates on the full system will fail for thermal states, and likewise will fail for eigenstates that have been reduced to subsystems.  Is there a way to recover $H$ using only the local correlations encoded by $M^{(\rho)}$, without direct knowledge of $\rho$?  In principle, the Gibbs state $\rho$ is fully determined by the 1-point expectation values $\Tr(\rho L_i)$ \cite{Swingle}.  But the procedure in \cite{Swingle} for reconstructing $\rho$ from the values $\Tr(\rho L_i)$ may have computational complexity that is exponential in the system size. One alternative is to examine $\bar{M}^{(\rho)}$ as defined in Equation \ref{eq:corrMatMixedComm}.  Indeed, numerical tests demonstrate that the kernel of $\bar{M}^{(\rho)}$ is generically $\ket{H}$, but the procedure appears highly sensitive to error, and the entries of $\bar{M}^{(\rho)}$ are not expectation values given by simple measurements, unlike the entries of $M^{v}$ for a pure state $v$. \footnote{One promising approach would be to make use of efficient (polynomial time) algorithms for generating tensor-network representations for Gibbs states of local Hamiltonians \cite{Molnar}. To recover $H$ from $M^{(\rho)}$, perhaps one could efficiently search over all local Hamiltonians for the Hamiltonian whose Gibbs state yields the desired correlation functions.}

For the case of a translation-invariant Hamiltonian, the approximate recovery process is very simple.   Let $v$ be the eigenstate of the total system of size $n$, and consider a singly connected subsystem $A$ of size $m$.  Then the $q=0$ part of $M^{(v)}$, or $\tilde M^{(v)}_{0\alpha,0\beta}$, may be estimated using $M^{(\rho_A)}$ with error that decays with $m$ as $e^{-m/\xi}$ for correlation length $\xi$. Using the estimate of $M^{(v)}_{0\alpha,0\beta}$ obtained from $M^{(\rho_A)}$, one can then find the kernel, yielding the Hamiltonian.

For the case of a non-translation-invariant Hamiltonian with an eigenstate that exhibits exponential decay of correlations, the ability to recover the Hamiltonian from a sub-region is less clear, and we are unable to comment conclusively.  One might expect that the eigen-operator of $M^{(\rho_A)}$ with the smallest eigenvalue approximately recovers the Hamiltonian $H_A$, at least if the terms near the boundary of $A$ are ignored when comparing the original Hamiltonian to the reconstruction.  In numerical examples for spin chains of length $n=12$ and sub-regions of length $n=8$, the Hamiltonian reconstructed from the eigenstate of the sub-region matched the Hamiltonian on region, ignoring boundary terms, with an accuracy of $\theta \approx 20^{\circ}-40^{\circ}$ using $\theta$ as defined in Equation \ref{eq:theta}.

\subsection{Unique reconstruction in the large-system limit}\label{sec:largeLimit}

Numerical examples readily demonstrate that the Hamiltonian may be uniquely recovered from an eigenstate for finite-size systems, and the arguments of Section \ref{sec:genericity} show that even one such numerical example is sufficient to make rigorous statements about almost all local Hamiltonians on finite-size systems.  We expect, moreover, that such examples exist for arbitrarily large finite-size systems -- perhaps they could be found analytically, for instance, with exactly solvable models.  In that case, one could conclude that almost all local Hamiltonians are uniquely determined by a single eigenstate, even for arbitrarily large systems.  However, while this statement is likely true, it ignores two important and related questions: the question of sensitivity to error, and the question of the infinite-size limit.  

Consider the value of $\lambda_2$ as a function on the space of local Hamiltonians, $LH$.  Recall from Section \ref{sec:sensitivity} that wherever $\lambda_2$ is nonzero, the Hamiltonian is uniquely determined by the eigenstate, and the sensitivity of the reconstruction to error goes like $(\lambda_2)^{-1}$.  One might worry about the behavior of this function as the system size increases: even though $\lambda_2$ may remain almost everywhere nonzero for large systems, does it become arbitrarily small in some regions of $LH$?  And for a formally infinite system, is the Hamiltonian still uniquely determined by the eigenstate?  

For translation-invariant systems, at least, if one has shown the ability to approximately reconstruct the Hamiltonian from the eigenstate on a sub-region, then these concerns become irrelevant: as system size increases, the reduced density matrix of the ground state on a fixed sub-region converges, at least for typical gapped ground states \cite{Cho}, and the Hamiltonian may always be reconstructed from this sub-region.  Using results from \cite{Cho}, combined with those of Section \ref{sec:subregion}, one could possibly use a numerical example of some Hamiltonian on a finite-size system to lower bound $\lambda_2$ for the same Hamiltonian on all larger systems, thus proving the Hamiltonian is recoverable with some bounded error on systems of arbitrary size.

For non-translation-invariant systems, recovery in the large-system limit remains unclear, which we leave as the subject of future work.  And even in translation-invariant systems, for spatial dimension $d \geq 2$, little is known rigorously about the ground states of generic local Hamiltonians, so it is difficult to speculate about the zeros of the correlation spectrum in different phases. 

\section{Conclusion and discussion}

In summary, we found that for finite-size lattice systems, a generic local Hamiltonian may be recovered from a single eigenstate.  To reconstruct the Hamiltonian from the eigenstate, we made use of the correlation matrix of the state, which encodes all connected two-point correlations between range-$k$ local observables.  The kernel of the correlation matrix yields the reconstructed Hamiltonian, and the reconstruction is possible within polynomial time, demanding knowledge only of two-point correlation functions of range-$k$ observables. The correlation spectrum indicates whether unique reconstruction of the Hamiltonian is possible, as well as the sensitivity of the reconstruction to error.  For translation-invariant systems, the correlation spectrum may be arranged in momentum bands, which appear to exhibit different structure for ground states and excited states.    Finally, in some cases it is possible to reconstruct the Hamiltonian knowing only the correlations of the state within a restricted region.  However, more work is necessary in this direction.

The correlation spectrum may be of independent interest for analyzing the correlation properties of a state.  On one hand, one may think the correlation spectrum captures little information that is not already captured by a standard quantity, like a correlation length or entanglement entropy.  However, unlike the correlation length of a particular observable, the correlation spectrum (and corresponding eigen-operators) offers canonically packaged data that is independent of any particular choice of observable. The correlation spectrum for a general state may be an analog of the scaling dimension spectrum in a conformal field theory, while the eigen-operators are an analog of conformal quasiprimary operators. The correlation spectrum also provides more numerical data than a single number like entanglement entropy, and one may investigate the $k$-correlation spectrum for different $k$.  Finally, the average correlation spectrum may be analytically calculable for certain ensembles of random local matrices.

\section*{Acknowledgments}
We would like to thank Tarun Grover, David Huse, and Geoff Penington for useful comments.  This work is supported by the David and Lucile Packard Foundation, and by the National Science Foundation through the grant No. PHY-1720504.

\newpage

\bibliography{references} 

\begin{thebibliography}{10}

\bibitem{Grover}
J.~R. Garrison and T.~Grover, ``Does a single eigenstate encode the full
  hamiltonian d?,'' {\em Physical Review X}, vol.~8, no.~2, p.~021026, 2018.
\newblock \url{https://doi.org/10.1103/PhysRevX.8.021026}.

\bibitem{Deutsch}
J.~M. Deutsch, ``Quantum statistical mechanics in a closed system,'' {\em
  Physical Review A}, vol.~43, no.~4, p.~2046, 1991.
\newblock \url{https://doi.org/10.1103/PhysRevA.43.2046}.

\bibitem{Srednicki}
M.~Srednicki, ``Chaos and quantum thermalization,'' {\em Physical Review E},
  vol.~50, no.~2, p.~888, 1994.
\newblock \url{https://doi.org/10.1103/PhysRevE.50.888}.

\bibitem{Osborne}
H.~L. Haselgrove, M.~A. Nielsen, and T.~J. Osborne, ``Quantum states far from
  the energy eigenstates of any local hamiltonian,'' {\em Physical review
  letters}, vol.~91, no.~21, p.~210401, 2003.
\newblock \url{https://doi.org/10.1103/PhysRevLett.91.210401}.

\bibitem{Zeng}
J.~Chen, Z.~Ji, Z.~Wei, and B.~Zeng, ``Correlations in excited states of local
  hamiltonians,'' {\em Phys. Rev. A}, vol.~85, p.~040303, Apr 2012.
\newblock \url{https://doi.org/10.1103/PhysRevA.85.040303}.

\bibitem{Cramer}
M.~Cramer, M.~B. Plenio, S.~T. Flammia, R.~Somma, D.~Gross, S.~D. Bartlett,
  O.~Landon-Cardinal, D.~Poulin, and Y.-K. Liu, ``Efficient quantum state
  tomography,'' {\em Nature communications}, vol.~1, p.~149, 2010.
\newblock \url{https://doi.org/10.1038/ncomms1147}.

\bibitem{Swingle}
B.~Swingle and I.~H. Kim, ``Reconstructing quantum states from local data,''
  {\em Physical review letters}, vol.~113, no.~26, p.~260501, 2014.
\newblock \url{https://doi.org/10.1103/PhysRevLett.113.260501}.

\bibitem{Verstraete}
V.~Zauner, D.~Draxler, L.~Vanderstraeten, J.~Haegeman, and F.~Verstraete,
  ``Symmetry breaking and the geometry of reduced density matrices,'' {\em New
  Journal of Physics}, vol.~18, no.~11, p.~113033, 2016.
\newblock \url{https://doi.org/10.1088/1367-2630/18/11/113033}.

\bibitem{Huse}
H.~Kim, M.~C. Ba{\~n}uls, J.~I. Cirac, M.~B. Hastings, and D.~A. Huse,
  ``Slowest local operators in quantum spin chains,'' {\em Physical Review E},
  vol.~92, no.~1, p.~012128, 2015.
\newblock \url{https://doi.org/10.1103/PhysRevE.92.012128}.

\bibitem{Vidal}
T.~O'Brien, D.~A. Abanin, G.~Vidal, and Z.~Papi{\'c}, ``Explicit construction
  of local conserved operators in disordered many-body systems,'' {\em Physical
  Review B}, vol.~94, no.~14, p.~144208, 2016.
\newblock \url{https://doi.org/10.1103/PhysRevB.94.144208}.

\bibitem{Zanardi}
P.~Zanardi, P.~Giorda, and M.~Cozzini, ``Information-theoretic differential
  geometry of quantum phase transitions,'' {\em Physical review letters},
  vol.~99, no.~10, p.~100603, 2007.
\newblock \url{https://doi.org/10.1103/PhysRevLett.99.100603}.

\bibitem{DavisKahan}
C.~Davis and W.~M. Kahan, ``The rotation of eigenvectors by a perturbation.
  iii,'' {\em SIAM Journal on Numerical Analysis}, vol.~7, no.~1, pp.~1--46,
  1970.
\newblock \url{https://doi.org/10.1137/0707001}.

\bibitem{Molnar}
A.~Molnar, N.~Schuch, F.~Verstraete, and J.~I. Cirac, ``Approximating gibbs
  states of local hamiltonians efficiently with projected entangled pair
  states,'' {\em Physical review b}, vol.~91, no.~4, p.~045138, 2015.
\newblock \url{https://doi.org/10.1103/PhysRevB.91.045138}.

\bibitem{Cho}
J.~Cho, ``Correlations in quantum spin systems from the boundary effect,'' {\em
  New Journal of Physics}, vol.~17, no.~5, p.~053021, 2015.
\newblock \url{https://doi.org/10.1088/1367-2630/17/5/053021}.

\end{thebibliography}
\bibliographystyle{ieeetr}

\end{document}